\documentstyle[multicol,aps,psfig]{revtex}
\renewcommand{\narrowtext}{\begin{multicols}{2}
\global\columnwidth20.5pc\noindent}
\renewcommand{\widetext}{\end{multicols}
\global\columnwidth42.5pc}
\begin{document}
\draft
\preprint{30 August 2004}
\title{Photoinduced Absorption Spectra of Halogen-Bridged Binuclear Metal
       Complexes:\\
       Possible Contrast between
       $\mbox{\boldmath$R$}_4$[Pt$_2$(P$_2$O$_5$H$_2$)$_4
        \mbox{\boldmath$X$}$]$\mbox{\boldmath$\cdot n$}$H$_2$O and
       Pt$_2$(CH$_3$CS$_2$)$_4$I}
\author{Jun Ohara and Shoji Yamamoto}
\address{Division of Physics, Hokkaido University,
         Sapporo 060-0810, Japan}
\date{Received 30 August 2004}
\maketitle
\begin{abstract}
The optical conductivity of photogenerated solitons in
quasi-one-dimensional halogen-bridged binuclear metal ($M\!M\!X$)
complexes is investigated with particular emphasis on a comparison
between the two family compounds
 $R_4$[Pt$_2$(pop)$_4X$]$\cdot$$n$H$_2$O
($X=\mbox{Cl},\mbox{Br},\mbox{I}$;
 $R=\mbox{NH}_4,\mbox{Na},\mbox{K},\cdots$;
 $\mbox{pop}=\mbox{diphosphonate}
 =\mbox{P}_2\mbox{O}_5\mbox{H}_2^{\,2-}$)
 and
 Pt$_2$(dta)$_4$I
($\mbox{dta}=\mbox{dithioacetate}=\mbox{CH}_3\mbox{CS}_2^{\,-}$).
Soliton-induced absorption spectra for the pop complexes should split into
two bands, while those for the dta complex should consist of a single
band.
\end{abstract}
\narrowtext

   Quasi-one-dimensional halogen ($X$)-bridged metal ($M$) complexes
\cite{G6408,W6435}, which are referred to as $M\!X$ chains, provide an
interesting stage \cite{N3865,N302,N427} performed by electron-electron
correlation, electron-lattice interaction and low dimensionality.
The Mott and Peierls insulators compete with each other in their ground
states \cite{Y422,Y2683}, while their photoexcited states exhibit novel
decay kinetics \cite{M5758,M5763}.
A large choice of metals, bridging halogens, ligand molecules and counter
ions enables us to investigate electron-phonon cooperative phenomena in
the one-dimensional Peierls-Hubbard system  systematically \cite{O2023}.

   In recent years, binuclear metal analogs which are referred to as
$M\!M\!X$ chains have stimulated renewed interest in this system,
exhibiting a wider variety of ground states
\cite{K533,K47,K1545,K435,Y183,Y13,Y125124}.
The existent $M\!M\!X$ compounds consist of two families:
 $R_4$[Pt$_2$(pop)$_4X$]$\cdot$$n$H$_2$O
($X=\mbox{Cl},\mbox{Br},\mbox{I}$;
 $R=\mbox{NH}_4,\mbox{Na},\mbox{K},\cdots$;
 $\mbox{pop}=\mbox{diphosphonate}
 =\mbox{P}_2\mbox{O}_5\mbox{H}_2^{\,2-}$) \cite{C4604,C409}
and
 $M_2$(dta)$_4$I
($M=\mbox{Pt},\mbox{Ni}$;
 $\mbox{dta}=\mbox{dithioacetate}
 =\mbox{CH}_3\mbox{CS}_2^{\,-}$) \cite{B444,B2815}.
The former compounds structurally resemble conventional $M\!X$ ones
and generally exhibit the same type of mixed-valent ground state
with halogen-sublattice dimerization \cite{K40,B1155}:
$-X^{-}\!\cdots\mbox{Pt}^{2+}\mbox{Pt}^{2+}\!\cdots
 X^{-}\!-\mbox{Pt}^{3+}\mbox{Pt}^{3+}\!-X^{-}\!\cdots$,
which is referred to as the charge-density-wave (CDW) state.
Pt$_2$(dta)$_4$I exhibits a distinct ground state with metal-sublattice
dimerization \cite{K10068}:
$\cdots\mbox{I}^{-}\!\cdots\mbox{Pt}^{2+}\mbox{Pt}^{3+}\!-\mbox{I}^{-}
 \!-\!\mbox{Pt}^{3+}\mbox{Pt}^{2+}\!\cdots\mbox{I}^{-}\!\cdots$,
which is referred to as the alternate charge-polarization (ACP) state.
Ni$_2$(dta)$_4$I is a Mott insulator \cite{S265} and
has a mono-valent ground state without any lattice distortion.
These ground states can be tuned by pressure
\cite{S1405,S66,Y140102,M046401,I2149} as well as by replacing the
bridging halogens \cite{B1155,Y1198}, counter ions
\cite{M046401,Y2321,M101} and ligand molecules \cite{I387}.

   Topological excitations of such competing ground states must provide
rich physics.
Soliton \cite{Y189} and polaron \cite{Y165113} solutions  have indeed been
found for an $M\!M\!X$ Hamiltonian of the Su-Schrieffer-Heeger type
\cite{S1698}, and an analogy between $M\!M\!X$ chains and
{\it trans}-polyacetylene has been pointed out.
The solitons turned out to have lower formation energies and smaller
effective masses than the polarons.
The direct $M(d_{z^2})$-$M(d_{z^2})$ overlap effectively reduces the
on-site Coulomb repulsion, and therefore, electrons can be more itinerant
in $M\!M\!X$ chains.
In fact $M\!M\!X$ chains exhibit a much higher room-temperature
conductivity than $M\!X$ chains \cite{K1931}.
Then we take more and more interest in solitons as charge or spin
carriers.
Thus motivated, we study the optical conductivity of $M\!M\!X$ solitons
with particular emphasis on a comparison between the pop and dta
complexes.
$M\!M\!X$ uniform absorption spectra have recently been investigated
\cite{K2163}, but photoinduced ones, which are supposed to serve as
prominent probes for nonlinear excitations, have neither measured nor
calculated yet.
Let us start exploring photoexcited mixed-valent binuclear metal complexes.

   We describe $M\!M\!X$ chains in terms of the one-dimensional
$\frac{3}{4}$-filled single-band Peierls-Hubbard Hamiltonian
\begin{eqnarray}
   &&
   {\cal H}
   =-t_{M\!M}\sum_{n,s}
     \bigl(b_{n,s}^\dagger a_{n,s}+a_{n,s}^\dagger b_{n,s}\bigr)
   \nonumber \\
   &&\quad
    -\sum_{n,s}
     \bigl[t_{M\!X\!M}-\alpha(v_{n+1}-v_n)\bigr]
     \bigl(a_{n+1,s}^\dagger b_{n,s}+b_{n,s}^\dagger a_{n+1,s}\bigr)
   \nonumber \\
   &&\quad
    -\beta\sum_{n,s}
     \bigl[
      (v_n-u_{n-1})n_{n,s}+(u_n-v_n    )m_{n,s}
     \bigr]
   \nonumber \\
   &&\quad
    +U_{M}\sum_{n}(n_{n,+}n_{n,-}+m_{n,+}m_{n,-})
   \nonumber \\
   &&\quad
    +V_{M\!M}\sum_{n,s,s'}n_{n,s}m_{n,s'}
    +V_{M\!X\!M}\sum_{n,s,s'}n_{n+1,s}m_{n,s'}
   \nonumber \\
   &&\quad
   +\frac{K_{M\!X}}{2}\sum_{n}
    \bigl[(u_n-v_n)^2+(v_n-u_{n-1})^2\bigr],
   \label{E:H}
\end{eqnarray}
where
$n_{n,s}=a_{n,s}^\dagger a_{n,s}$ and
$m_{n,s}=b_{n,s}^\dagger b_{n,s}$ with
$a_{n,s}^\dagger$ and $b_{n,s}^\dagger$ being the creation
operators of an electron with spin $s=\pm$ (up and down) for the
$M\,d_{z^2}$ orbitals in the $n$th $M\!M\!X$ unit.
$t_{M\!M}$ and $t_{M\!X\!M}$ describe the intradimer and interdimer
electron hoppings, respectively.
$\alpha$ and $\beta$ are the intersite and intrasite electron-lattice
coupling constants, respectively, with $K_{M\!X}$ being the
metal-halogen spring constant.
$u_n$ and $v_n$ are, respectively, the chain-direction displacements
of the halogen and metal dimer in the $n$th $M\!M\!X$ unit from their
equilibrium positions.
We assume, based on the thus-far reported experimental observations,
that every $M_2$ moiety is not deformed.
We set $t_{M\!X\!M}$ and $K_{M\!X}$ both equal to unity in the following.

   Here, we focus our research on platinum complexes with mixed-valent
ground states.
The intradimer transfer integral and Coulomb interactions are, unless
otherwise noted, set for
$t_{M\!M}=2.0,U_{M}=1.0,V_{M\!M}=0.5,V_{M\!X\!M}=0.3$.
The Coulomb repulsion is much weaker on platinum
\vspace*{-4mm}
\begin{figure}
\centerline
{\mbox{\psfig{figure=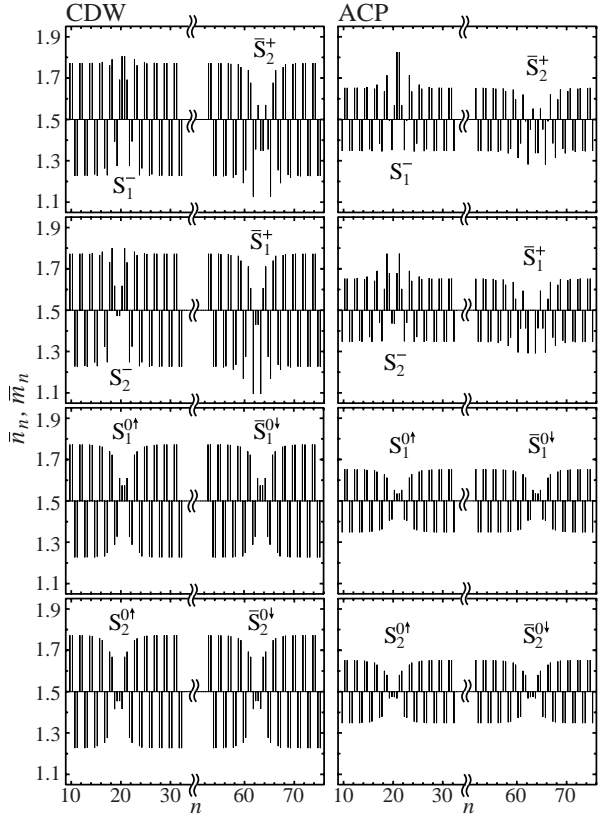,width=80.0mm,angle=0}}}
\caption{Spatial configurations of the stable soliton-antisoliton pairs on
         the CDW and ACP backgrounds,  where quantum averages of the local
         electron densities,
         $\sum_s\langle a_{n,s}^\dagger a_{n,s}\rangle\equiv\bar{n}_n$ and
         $\sum_s\langle b_{n,s}^\dagger b_{n,s}\rangle\equiv\bar{m}_n$,
         are measured in comparison with the average occupancy.}
\label{F:SS}
\end{figure}
\vspace*{-6mm}
\begin{figure}
\centerline
{\mbox{\psfig{figure=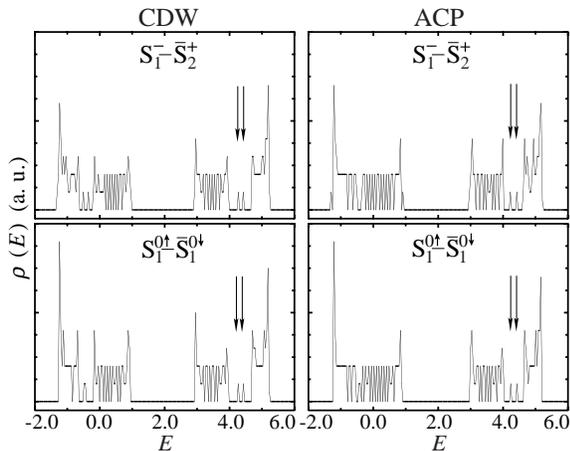,width=76.0mm,angle=0}}}
\caption{Density of states for the optimum soliton-antisoliton pairs on
         the CDW and ACP backgrounds.}
\label{F:DOS}
\end{figure}
\noindent
ions than on nickel ions.
The electron-lattice coupling constants are taken in two ways as
$\alpha=0.0,\beta=1.4$and $\alpha=0.3,\beta=1.0$, which are relevant to
the pop and dta complexes and indeed give the CDW and ACP ground states,
respectively, under the above parametrization.
The lattice distortion is adiabatically determined at each temperature so
as to minimize the free energy, where the chain length is kept unchanged.

   The optical spectra are obtained by calculating the matrix elements
between the ground state $|{\rm g}\rangle$ of energy $E_{\rm g}$ and the
excited states $|l\rangle$ of energy $E_l$ ($l=1,2,\cdots$) for the
current operator ${\cal J}=\sum_{n=1}^N\sum_{s=\pm}j_{n,s}$ with
\begin{eqnarray}
   &&
   j_{n,s}
   =\frac{{\rm i}e}{\hbar}
    c_{M\!M}t_{M\!M}
    \bigl(b_{n,s}^\dagger a_{n,s}-a_{n,s}^\dagger b_{n,s}\bigr)
   +\frac{{\rm i}e}{\hbar}
    c_{M\!X\!M}
   \nonumber\\
   &&\ \times
    \bigl[t_{M\!X\!M}
     -\alpha(v_{n+1}-v_n)\bigr]
     \bigl(a_{n+1,s}^\dagger b_{n,s}-b_{n,s}^\dagger a_{n+1,s}\bigr),
\end{eqnarray}
where $c_{M\!M}$ and $c_{M\!X\!M}$ are the average $M$-$M$ and $M$-$X$-$M$
distances, respectively, and are set for $c_{M\!X\!M}=2c_{M\!M}$.
The real part of the optical conductivity is given by
\begin{equation}
   \sigma(\omega)
    =\frac{\pi}{N\omega}\sum_l
   |\langle l|{\cal J}|{\rm g}\rangle|^2
   \delta(E_l-E_{\rm g}-\hbar\omega).
\end{equation}
$|{\rm g}\rangle$ is set for the Hartree-Fock (HF) ground state, while
$|l\rangle$ is calculated within and beyond the HF approximation, being
generally defined as
\begin{equation}
   |l\rangle
   =\sum_s
    \sum_{\epsilon_\mu\leq\epsilon_{\rm F}}
    \sum_{\epsilon_\nu>\epsilon_{\rm F}}
    f(\mu,\nu,s;l)c_{\nu,s}^\dagger c_{\mu,s}|{\rm g}\rangle,
\end{equation}
where $\epsilon_{\rm F}$ is the Fermi energy and $c_{\lambda,s}^\dagger$
creates an electron with spin $s$ for the $\lambda$th HF eigenstate with
an eigenvalue $\epsilon_\lambda$.
At the HF level, any excited state is simply approximated by a single
Slater determinant as $f(\mu,\nu,s;l)=\delta_{\mu\nu s,l}$.
In order to take account of the excitonic effect, we further consider
excited states of the configuration-interaction (CI) type, where
$f(\mu,\nu,s;l)$ is determined so as to diagonalize the original
Hamiltonian (\ref{E:H}).
We set $N$, the number of unit cells, equal to $84$, which results in
spending $2$ GB memory on the CI calculation.

   Photogenerated defects are necessarily in pairs.
We visualize the convergent soliton (S)-antisoliton ($\bar{\mbox{S}}$)
pairs in Fig. \ref{F:SS}, which have been calculated at a sufficiently low
temperature $k_{\rm B}T/t_{M\!X\!M}=10^{-3}$ without any assumption on
their shapes in an attempt to elucidate the intrinsic excitation
mechanism.
S$_1$ and S$_2$ designate the lowest- and highest-energy soliton
solutions, both of which lay their centers on halogen sites (metal dimers)
with the CDW (ACP) background.
There are further soliton solutions with intermediate energies, but they
are unstable in pairs.

\begin{figure}
\centerline
{\mbox{\psfig{figure=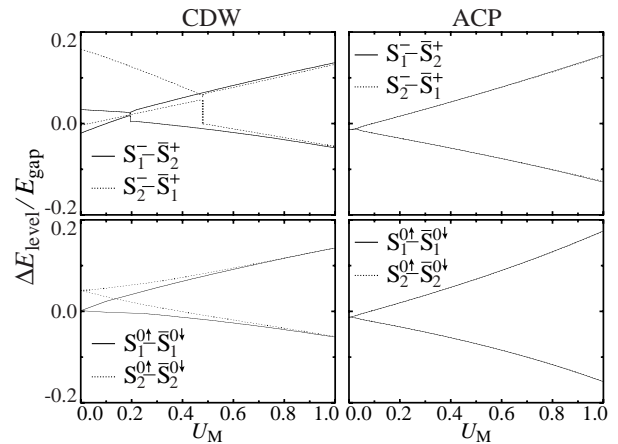,width=80.0mm,angle=0}}}
\caption{Energy shifts of the localized levels accompanying the stable
         soliton-antisoliton pairs on the CDW and ACP backgrounds, which
         are measured from the gap center and scaled by the Peierls gap,
         as functions of the Coulomb interaction, where $U_M$ varies
         keeping the relation $U_M=V_{M\!M}/0.5=V_{M\!X\!M}/0.3$.}
\label{F:level}
\end{figure}

\begin{figure}
\centerline
{\mbox{\psfig{figure=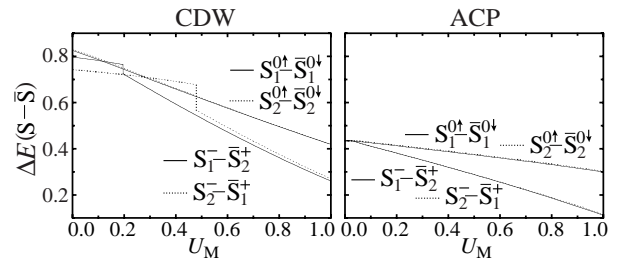,width=80.0mm,angle=0}}}
\caption{Formation energies of the stable soliton-antisoliton pairs
         on the CDW and ACP backgrounds as functions of the Coulomb
         interaction, where $U_M$ varies keeping the relation
         $U_M=V_{M\!M}/0.5=V_{M\!X\!M}/0.3$.}
\label{F:Ef}
\end{figure}

   Such a pair of solitons generally gives two additional levels within
the gap, which are indicated by arrows in Fig. \ref{F:DOS}.
The lower one is doubly filled, while the upper one is vacant.
There appear further soliton-related mid-gap levels in the strong-coupling
and/or -correlation region \cite{T4074,T1800,T2212}.
These levels are all localized around the soliton centers \cite{Y189}.
In Fig. \ref{F:DOS}, the lower and upper ones are assigned to $\mbox{S}^-$
and $\mbox{S}^+$, respectively, in the case of charged soliton pairs,
whereas both the levels originate from a single $\mbox{S}^0$ itself in the
case of neutral soliton pairs because of the breakdown of the spin up-down
symmetry.
The intragap soliton levels are plotted as functions of the Coulomb
interaction in Fig. \ref{F:level}.
The Coulomb effect distinguishes charged soliton pairs
$\mbox{S}^-$$-$$\bar{\mbox{S}}^+$ from singly excited charged solitons
$\mbox{S}^\pm$.
Without any Coulomb interaction, the two localized levels of
$\mbox{S}^-$$-$$\bar{\mbox{S}}^+$ are essentially the same as those of
single $\mbox{S}^\pm$.
In the case of the CDW background, with increasing Coulomb interaction,
the levels of $\mbox{S}^-$ and $\mbox{S}^+$ move upward and downward,
respectively, and then cross.
For further increasing Coulomb interaction, the doubly occupied
$\mbox{S}^-$ level goes higher in energy than the vacant $\mbox{S}^+$
level, provided that they are singly excited.
Therefore, in such a strong correlation regime,
$\mbox{S}^-$ and $\mbox{S}^+$ are restructured in a pair so as to have
lower- and higher-lying levels, respectively, where the pair-creation
energy $\Delta E(\mbox{S}^-$$-$$\bar{\mbox{S}}^+)$ is smaller than twice
the single-excitation energy $2\Delta E(\mbox{S}^\pm)$.
Hence, $\Delta E(\mbox{S}^-$$-$$\bar{\mbox{S}}^+)$ jumps at certain
Coulomb-interaction values, as shown in Fig. \ref{F:Ef}.
Since the same scenario applies in the case of increasing Peierls coupling
$\alpha$, $\mbox{S}^-$$-$$\bar{\mbox{S}}^+$ and single $\mbox{S}^\pm$
behave differently on the ACP background lying in the large-$\alpha$
region, regardless of Coulomb interaction.
Thus, photoinduced and doping-induced charged solitons should exhibit
distinct absorption spectra in general.
This is not the case with neutral solitons.
The level structure, formation energy and absorption spectrum of any
photoinduced $\mbox{S}^{0\uparrow}$$-$$\bar{\mbox{S}}^{0\downarrow}$ can
be obtained by superposing those of singly excited $\mbox{S}^{0\uparrow}$
and $\mbox{S}^{0\downarrow}$.
Photoinduced and chemical-defect-induced neutral-soliton absorption
spectra may essentially be the same.
\vspace*{-0.1mm}

\begin{figure}
\centerline
{\mbox{\psfig{figure=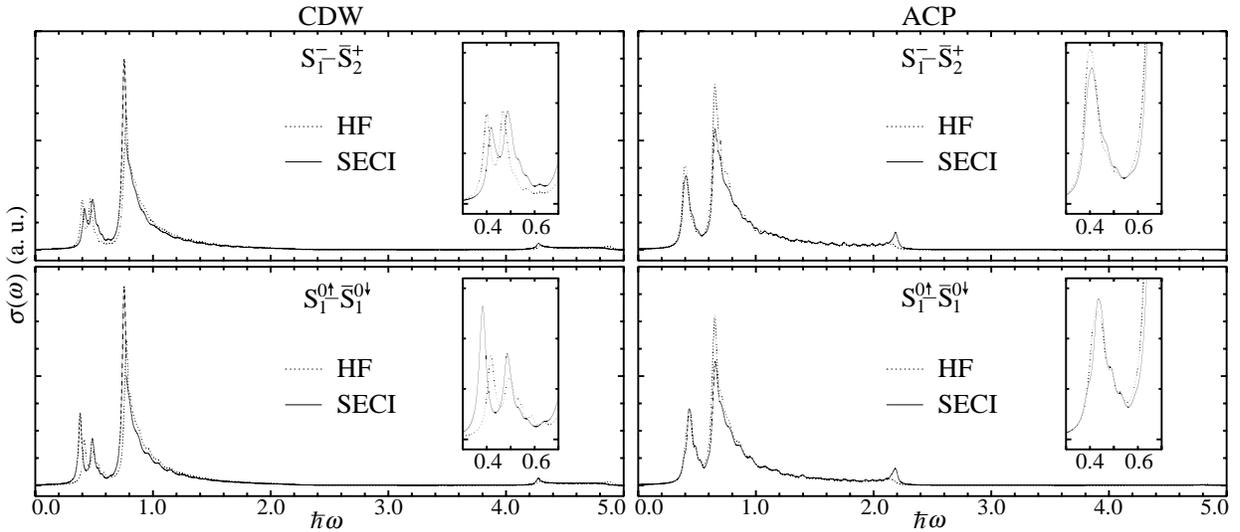,width=164mm,angle=0}}}
\vspace*{1mm}
\caption{Hartree-Fock (HF) and single-excitation configuration-interaction
         (SECI) calculations of the optical conductivity spectra for the
         optimum soliton-antisoliton pairs on the CDW and ACP backgrounds,
         where each line has been Lorentzian-broadened.
         Mid-gap absorption bands due to solitons are scaled up in
         insets.}
\label{F:OC}
\end{figure}

   Another Coulomb effect causes a striking contrast between the CDW and
ACP solitons.
With increasing Coulomb interaction, the intragap soliton levels generally
move away from the gap center.
Since the electron-hole symmetry is broken in the Hamiltonian (\ref{E:H}),
soliton-related electron and hole levels may be asymmetric with respect to
the center of the gap.
It is indeed the case with the CDW background, while on the ACP
background, a soliton of charge $\sigma$, S$^{\sigma}$, and that of spin
$s$, S$^{0s}$, described in terms of electrons are still almost equivalent
to their counterparts S$^{-\sigma}$ and S$^{0\,-s}$ described in terms of
holes, respectively.
In consequence, {\it photoinduced soliton absorption spectra for the pop
complexes should split into two bands, while those for Pt$_2$(dta)$_4$I
should consist of a single band}, as demonstrated in Fig. \ref{F:OC}.
In mixed-valent $M\!X$ chains, neutral solitons seem to be the
lowest-energy pair excitations \cite{I1088}.
Photoexcited [Pt(en)$_2X$](ClO$_4$)$_2$ ($X=\mbox{Cl},\mbox{Br}$;
 $\mbox{en}=\mbox{ethylenediamine}=\mbox{C}_2\mbox{H}_8\mbox{N}_2$)
\cite{K2237,O861} indeed exhibit mid-gap absorption attributable to
neutral solitons.
In mixed-valent $M\!M\!X$ chains, on the other hand, charged solitons may
be the lowest-energy excitations because the on-site Coulomb repulsion
$U_M$ and the Holstein coupling $\beta$ are effectively smaller and
larger, respectively \cite{Y125124,K10068}.
Polarons have much higher formation energies for both $M\!X$
\cite{G6408,I1088} and $M\!M\!X$ \cite{Y165113} chains and can therefore
be generated from relatively high-energy excited states corresponding to
the electron-hole continuum \cite{M5763,O2023}.
An excitation energy close to the Peierls gap directly induces
charge-transfer excitons \cite{M5758} and they may relax into soliton
pairs in nonradiative channels for $M\!M\!X$ chains as well.
Then there arises an interesting issue: charged solitons or neutral
solitons?
The optical conductivity spectra possibly answer this question.
For the pop complexes, charged (neutral)-soliton mid-gap absorption
spectra are double-peaked and the higher (lower)-energy band has a larger
oscillator strength, where the excitonic effect is essential.
As for Pt$_2$(dta)$_4$I, further experiments such as
electron-spin-resonance measurements \cite{T2169} are supplementary to
distinguish between charged and neutral solitons.

   The doublet structure of soliton-induced spectra is characteristic of
$M\!M\!X$ chains and is a consequence of {\it the broken electron-hole
symmetry and effective Coulomb correlation} \cite{O115112}.
Indeed $M\!X$ chains also lose the electron-hole
\vspace*{-3mm}
\vspace*{-5mm}

\widetext
\vspace*{1mm}
\narrowtext
\noindent
symmetry with their
$X$ $p_z$ electrons activated \cite{Y125124,G10566}, but it is not the
case with the Pt$X$ compounds, where the energy level of the $X$ $p_z$
orbitals is much lower than that of the $M$ $d_{z^2}$ orbitals, and thus,
the $p$-orbital contribution may effectively be incorporated into the
intermetal supertransfer energy of a single-band Hamiltonian of the
eq. (\ref{E:H}) type.
In fact, the photoexcited [Pt(en)$_2$I](ClO$_4$)$_2$ \cite{O6330} and
[Pt(en)$_2$Br](ClO$_4$)$_2$ \cite{O861} yield single-peaked mid-gap
absorption spectra attributable to charged and neutral solitons,
respectively, though the latter $\mbox{S}^0$ spectrum is accompanied by a
weak shoulder \cite{I1380}.
Figure \ref{F:OC} demonstrates that a breakdown of the electron-hole
symmetry does not necessarily lead to the doublet structure.
The double-peaked soliton absorption is peculiar to the pop complexes but
missing in Pt$_2$(dta)$_4$I.
Observations of such a contrast will strongly support effective Peierls
coupling in Pt$_2$(dta)$_4$I.
Distinguishable observations of charged and neutral solitons in the pop
complexes will contribute toward realizing photoswitched charge or spin
conduction.
Photoinduced infrared absorption measurements on $M\!M\!X$ compounds
are strongly encouraged.

   The authors are grateful to K. Iwano and Y. Shimoi for fruitful
discussions and helpful comments.
This work was supported by the Ministry of Education, Culture, Sports,
Science and Technology of Japan and the Iketani Science and Technology
Foundation.

\widetext
\end{document}